# Experimental Study of Remote Job Submission and Execution on LRM through Grid Computing Mechanisms

Harshadkumar B. Prajapati
Information Technology Department
Dharmsinh Desai University
Nadiad, INDIA
e-mail: prajapatihb.it@ddu.ac.in,
harshad.b.prajapati@gmail.com

Vipul A. Shah
Instrumentation & Control Engineering
Department
Dharmsinh Desai University
Nadiad, INDIA
e-mail: vashahin2010@gmail.com,
vashah.ic@ddu.ac.in

*Abstract*—Remote job submission and execution is fundamental requirement of distributed computing done using Cluster computing. However, Cluster computing limits usage within a single organization. Grid computing environment can allow use of resources for remote job execution that are available in other organizations. This paper discusses concepts of batch-job execution using LRM and using Grid. The paper discusses two ways of preparing test Grid computing environment that we use for experimental testing of concepts. This paper presents experimental testing of remote job submission and execution mechanisms through LRM specific way and Grid computing ways. Moreover, the paper also discusses various problems faced while working with Grid computing environment and discusses their trouble-shootings. The understanding and experimental testing presented in this paper would become very useful to researchers who are new to the field of job management in Grid.

*Keywords-LRM; Grid computing; remote job submission; remote job execution; Globus; Condor; Grid testbed.*

I. INTRODUCTION

Research efforts in Grid computing [1] started due to limitation faced in the number of resources offered by Cluster computing [2]. Cluster computing contains homogeneous resources, which are under a single administrative control, whereas Grid computing can allow use of resources present across boundary of an organization [3]. Therefore, use of Grid computing becomes essential when resources of a single organization are not sufficient to solve their computation-demanding problem, generally, a scientific experiment. Grid computing technology and software can aggregate resources of various organization in collaborative way with considering various heterogeneity present in terms of architecture and software used at Cluster computing level.

The fundamental unit of work-done in Grid computing is successful execution of a job. A job in Grid is generally a batch-job, which does not interact with user while it is running. Higher-level Grid services and applications use job submission and execution facilities that are supported by a Grid middleware. Therefore, it is very important to understand how a job is submitted, executed, and monitored in Grid computing environment. Researchers and scientists who are working in higher-level Grid services and applications do not pay much attention to the underlying Grid infrastructure in their published work. However, such knowledge is very useful to beginners. Moreover, understanding of internal working of job submission and execution becomes very useful, in trouble-shooting, when the higher-level services or applications report failures.

In this research paper, we concisely present important concepts of Local Resource Manager and Grid computing environment. We discuss about preparing required Grid computing environment without spending much efforts and doing rework on real computing machines, for which we discuss about Grid environment made of Virtual Machines and made of real machines. Moreover, through experiments performed on our prepared Grid environment, we show direct remote job submission on LRM and through Grid computing mechanisms. The presented experimental study of remote job submission and execution on LRM through Grid computing mechanism becomes very useful in learning concepts practically and also when trouble-shooting underlying mechanisms of Grid environment.

The work in *Introduction to Grid Computing with Globus* [4] discusses installation steps of building Grid infrastructure using GT 4. A recent work in [5] focuses on implementation steps of achieving security in Grid through Grid Security Infrastructure (GSI). Specifically, their work [5] focuses on installation and testing of host certificates and client certificates and their testing. As compared to [5], our work has wider scope, not just security in Grid. As compared to [4], we focus on providing practical understanding of job execution on LRM using LRM mechanisms and two Grid computing mechanisms. Moreover, we also present an easy way of preparing Grid testbed without use of any networking hardware, which researchers can implement in their Personal Computer or Laptop. Furthermore, our work also focuses on trouble-shooting of important problems faced during use of Grid computing environment.

The rest of the paper is structured as follows. Section II concisely describes Local Resource Manager and its responsibilities and discusses why there is need of Grid mechanism to bridge heterogeneity of different LRMs. Section III describes Grid computing environment and its responsibilities and provided services. Section IV presents how to prepare Grid computing environment for practical study. Section V provides experimental testing of LRM using LRM mechanisms and discusses need of Grid computing mechanism. Section VI provides experimental testing of remote job submission and execution on LRM using Grid computing mechanisms. Section VII discusses major problems faced during use of Grid computing environment and also shows their trouble-shooting. Finally, Section VIII provides conclusions and future work.

## II. LOCAL RESOURCE MANAGER

In most Grid deployments, individual resources of a single organization are centrally managed using batch queue oriented local resource management system (LRM). In such system, user submitted batch jobs are introduced into the queue of LRM, from which resource scheduler decides about execution of jobs. Examples of such LRMs include PBS [6], LSF [7], Condor [8], Sun Grid Engine [9], and Load Leveler [10]. A resource scheduler is also known as a low-level scheduler or local scheduler, as it deals with the jobs submitted in a single administrative domain. For compute resources, LRM can be configured for (i) which user is allowed to run jobs, (ii) what policies are associated with selection of jobs for running, and (iii) what policies are associated with individual machines for considering them to be idle.

An LRM manages two types of jobs: local jobs generated inside a resource domain and jobs generated by external users, i.e. Grid users. The common purpose of any LRM is to manage, control, and schedule batch processes on the resources under its control. Since, LRM deals with distributed resources, it is also called Distributed Resource Manager. Basic features of any LRM include following:
- Scripting language for defining batch job
- Interfaces for submission of batch jobs
- Interfaces to monitor the executions
- Interfaces to submit input and gather output data
- Mechanism for defining priorities for jobs
- Match-making of jobs with resources
- Scheduling jobs present in queue(s) to determine execution order of jobs based on job-priority, resource status, and resource allocation configuration.

Advance reservation based LRMs are also used in Grid. Advance reservation allows reserving resources in advance for their use in future. Examples of such LRMs include Platform LSF [7], PBS Pro/Torque [11], [12], Maui [13], and SGE [9].

Grid allows use of heterogeneous LRMs through a common interface. In Globus based Grid, which is widely used, GRAM (Globus Resource Allocation Manager) interface is used for job submission on heterogeneous LRMs. GRAM is a standardized way of accessing any LRM, as it is LRM neutral. GRAM messages remain same irrespective of LRM, whether LRM is PBS, LSF, or Condor. Each LRM can be connected into a Globus based Grid using Globus-LRM adapter. Clients submit job requests using GRAM protocol and GRAM-LRM connector can translate those messages into a language understood by a specific LRM. For example, GRAM5 supports Condor, PBS, and LSF; similarly, Unicore [14] supports SGE, LoadLeveler and Torque.

We use HTCondor as LRM in our test Grid computing environment. Therefore, we concisely discuss about it. HTCondor [15], which is formerly known as Condor [8], is a set of daemons and commands that enable to implement concept of batch-queue controlled Cluster computing [2] of jobs. HTCondor is available for various platforms including Unix, Linux, and Windows OSes. We use HTCondor and Condor interchangeably to refer to the same thing.

The main goal of HTCondor is to provide high throughput distributed computing infrastructure that transparently executes a large number of jobs over a long period. Though, HTCondor allows use of computing power of a computing machine to users, it is the owner of the computing machine that defines under which conditions Condor can allow use of the computing machine to batch jobs. For example, HTCondor can be configured to run a job when keyboard and CPU are idle. Condor implements ClassAds (Class Advertisements) to describe requirements of jobs and specification of computing resources. Computing resources advertise their resource properties such as available memory, CPU type, CPU speed, virtual memory size, physical location, and current load average. Interaction with Condor system for various activities is done through command interface. While a job is under execution, input/output files of the job can be transferred using Condor I/O, which redirects I/O related system calls of a job to the job submit machine.

## III. GRID COMPUTING ENVIRONMENT

Grid computing enables utilization of idle time of resources that are available at geographically diverse locations [14], [16], [17]. Grid computing can allow access of resources such as storage, sensors, application software/code, databases, and computing power. In Grid, resources are autonomous and heterogeneous. Current Grid computing, in most deployments, is on collaborative manner, in which resources exhibit varying availability. However, QoS oriented Grid computing [18] is also possible. Grid computing has been used in drug discovery, GIS processing, sky image processing, industrial research, and scientific experiments. Use of Grid computing infrastructure can be done by converting traditional applications into Grid applications, which have form of either independent tasks application or dependent tasks application. The independent tasks applications include Parameter Sweep and Task farming (embarrassingly parallel) problems [19], e.g. drug discovery. Dependent tasks applications include workflow [20] applications, e.g., Montage [21].

Grid computing field started as an extension of Cluster computing. Both Cluster computing and Grid computing are used to utilize idle computing resources; however, there are differences between two. If Grid computing is compared with cluster computing, cluster computing has resources under central control and cannot connect resources that are present across boundary of the organization. Whereas, using Grid computing it is possible to connect, share, and use resources that are present in other organizations. Moreover, Grid allows heterogeneous resources, whereas Cluster does not. To enable use of diverse, idle resources, Grid computing field started in mid 90's with the term Meta-Computing. Two projects: Globus Project [22], at Argonne National Laboratories and the University of Chicago, and Legion [23], [24], at University of Virginia, are considered to be starting points for Grid computing field.

To implement Grid computing, we need Grid middleware. Grid middleware is a software that glues different local resources of organization(s) to form a higher-level, bigger, global resource. A Grid middleware offers following services:
- Submission and monitoring of remote jobs
- Allocation and co-allocation of resources
- Access to storage devices and data management
- Information service (kind of yellow page service)
- Resource discovery, resource registration, resource information update
- Security service: authentication, authorization, and delegation

Most Grid deployments are based on Globus, which is a de-facto middleware software for implementing Grid computing. Globus toolkit 5.2.3 (GT 5) [25], which we have used for experimentation in this paper, offers following components.
- Grid Security [26] (Certificate based authentication, authorization and single signon) and Simple CA certificate authority for signing host and user certificates.
- GSI (Certificate) based GridFTP [27] (client and server) for transferring data (files) on Grid sites (machines)
- GRAM client and GRAM server to allow remote job submission and execution using standard protocol (GRAM).

## IV. PREPARATION OF TEST ENVIRONMENT

Figure 1 shows architecture of Grid computing environment with focus on LRM and connecting them using Grid computing mechanism. A Grid computing environment involves various Grid sites, generally one organization is considered as one Grid site. Each organization generally contains batch-queue controlled cluster, which is exposed to external organizations through Grid computing services and protocols. Each Grid site contains a Head node, which is accessible through network, generally Internet. For implementation of Grid environment, we prepare three Grid sites in virtual Grid computing environment inside Oracle VirtualBox and prepare three Grid sites in real, physical Grid computing environment made of three computers connected through a network switch.

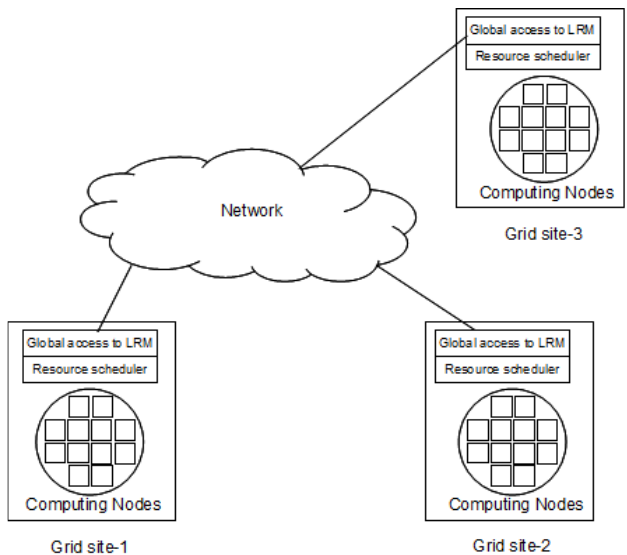

Figure 1. Architecture of connecting LRMs using Grid computing.

Initially, we installed Globus and other software on Virtual Machines prepared using Oracle VirtualBox software. Figure 2 shows three Virtual Machines of test Grid testbed. In virtual environment, we need to install Ubuntu OS only once on virtual machines (VMs), and then other Ubuntu machines can be created by simply cloning existing one. Cloning of virtual machine takes around 5 to 7 minutes; where as installation of Ubuntu OS on physical machine can take 30 to 45 minutes for installation.

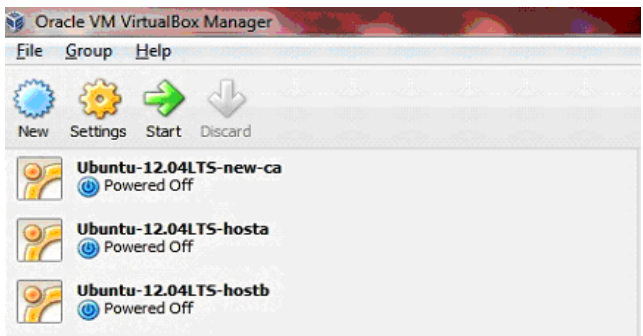

Figure 2. Screen-shot showing few constituent virtual machines of test Grid environment.

We followed VM installation approach first to face all problems before installation on physical machines, trouble-shoot and solve all installation problems before we deploy Grid software on actual machines. Using Virtual environment, we can do even network related activities without having physical network. In a single computer, we can create many virtual machines working as Grid nodes; however, we can have only few in running state. Available main memory limits the number of virtual machine instances we can keep running without crashing host Operating System. In our virtual implementation of Grid computing environment, we use a Laptop with 3 GB RAM and Windows 7 as operating system as a Host machine. Inside this host machine, we create three Ubuntu machines, each with 512 RAM, as virtual resources of Grid computing environment. We experimentally tested that our Host machine does not allow crash-free running of more than three virtual Grid nodes.

Preparation of physical test environment includes installation and configuration of following components:
- Ubuntu 12.04LTS [28]: As Globus works on Linux OS, we use Ubuntu 12.04LTS
- NTP [29]: Network Time Protocol is needed for synchronization of Clock
- HTCondor: We use it as Local Resource Manager.
- GSI: Grid Security Infrastructure is needed for X.509 certificate based authentication of Grid user
- GridFTP: It is GSI enabled File Transfer Protocol. It is used to transfer files.
- GRAM: Grid/Globus Resource Allocation Manager is used to manage job on Grid site. It is independent of LRMs
- Globus-Condor Adaptor: It is needed to convert GRAM protocol message into LRM specific messages.
- Condor-G: It enables Condor feature for Grid jobs. It enables Condor to use other LRMs

Following Globus Components are needed on machines that work as Grid Node.
- GridFTP (GridFTP server and client)
- Globus Gate Keeper (GRAM server and client)
- GSI configuration

Following Globus Components are needed on machine that works as Certificate Authority (CA).
- GSI configuration
- SimpleCA

In our physical Grid computing environment, we have three dual core machines: grid-b, grid-v, and ca, each with 1 GB RAM and Ubuntu 12.04LTS Operating System. For network connection, we use 8 port switch. The grid-b and grid-v machines work as Grid nodes and the ca machine works as Certificate Authority and also as Grid node. Moreover, each Grid node has its own LRM, i.e., ca grid-b, and grid-v machine represent separate Condor pools having their own Central Manager component of Condor. Therefore, ca machine can be seen as one Grid site, grid-b machine can be seen as another Grid site, and grid-v can be seen as another Grid site. It is possible to add additional computing nodes under each Grid site. To do that, we do not need to install Globus software on additional machines. However, we need to install LRM software, i.e., Condor in our test environment, on additional machines, and configure additional machines to report to particular Central Manager, i.e., either ca or grid-b or grid-v in our test environment.

V. EXPERIMENTAL TESTING OF JOB SUBMISSION AND EXECUTION ON LRM

Testing of working of Condor as a LRM is done by submitting a job to it.

Figure 3 shows how to check status of Condor pool.

Figure 3. Check status of Condor pool using condor_status

Figure 4 shows how to prepare a test submit file. A Condor job file, i.e., submit file, includes name of executable, name of output file, name of error file, and name of log file. The submit file is ended with Queue command, which tells to Condor system to queue the specified job. Specified executable and input file, in our example, input file is not shown and not required for /bin/hostname command, are transferred to the remote executing machine. Log file and output file are created in current directory on submit machine. For considered job in Figure 4, the output file result.out will contain output of /bin/hostname command.

```
$cat myjob.sub
Universe = Vanilla
Executable = /bin/hostname
Output = result.out
Error = result.err
Log = result.log
Queue
```

Figure 4. Condor submit file for its testing

Figure 5 shows how to submit a job using condor_submit and how to see status of completed job using condor_history.

Figure 5. Status of completed jobs using condor_history

When we work with any LRM, we need to use LRM specific commands to carry out various operations related to job submission and execution. In our experiments, we use Condor as LRM. In Torque LRM, qstat command is used to see status of a Torque cluster, and qsub command is used to submit a job to the cluster. In SGE, qhost command is used to see status of a SGE cluster, and qsub command is used to submit a job to the cluster. Moreover, user needs to prepare a job-submission script file, whose syntax varies depending upon used LRM. For example, job-submission script file of Condor would not work for SGE or Torque. This heterogeneity of ways of accessing different LRMs is handled by Grid computing architecture using GRAM servers and GRAM clients, which use GRAM protocol rather than LRM specific syntax.

VI. EXPERIMENTAL TESTING OF REMOTE JOB SUBMISSION AND EXECUTION ON LRM USING GRID COMPUTING MECHANISMS

We demonstrate two ways of performing remote job execution in Grid computing environment.

A. *Remote Job Submission to LRM using globus*

We show an example of testing of Condor LRM for remote job submission by running a job on ca machine from grid-b machine. It involves following steps.

**Step 1**: Login as Grid user and create proxy credential (see Figure 6). We are working on the machine: grid-b.it2.ddu.ac.in as gtuser user.

Figure 6. Create proxy credential using grid-proxy-init.

**Step 2**: Submit a job: We use /bin/date as a test job, and submit job to Condor on the remote machine: ca.it2.ddu.ac.in. We need to wait for some time until our submitted job gets executed on remote machine and output gets received on our submit computer (see Figure 7).

Figure 7. Submit a job using globus-job-run.

**Step 3**: Check Job entry and detail on ca.it2.ddu.ac.in for our submitted Job. The entry of above submitted job is with ID = 5.0 (see Figure 8).

Figure 8. Check job status using condor_history on ca machine.

## B. Remote Job Submission to LRM using Condor-g

The Condor-G allows Grid universe Job to be submitted using condor_submit command. We perform following steps to test remote job submission and execution.

**Step 1**: Prepare a test submit file (see Figure 9) to be used as submit file to Condor-G. We need to use Globus universe for Condor-G submission. Same Condor-G submission file can also be used to access other LRM. For example, if we want to submit the job to SGE LRM, then we need to use jobmanager-sge, on line no. 3, rather than jobmanager-condor, which is for Condor LRM.

```
$cat condortest.submit
Universe = Globus
grid_resource = gt5 grid-v.it2.ddu.ac.in/jobmanager-condor
Executable = /bin/hostname
Arguments = -f
Output = result.output
Error = result.error
Log = result.log
Queue
```

Figure 9. A test submit file to test working of Condor-G.

**Step 2**: Submit the job using condor_submit (see Figure 10). We get Job ID of the submitted Job as 9.

Figure 10. Submitting a condor-g job using condor_submit.

**Step 3**: Keep condor_q running on another terminal continuously using watch command. The job is in queue of condor_q and it is in idle state. Our job has ID = 9. In Figure 11, we can see state is I under ST column at time = 15:07:56.

Figure 11. Status of the job using condor_q. The job is idle.

In Figure 12, we can see Job is in running state. We can see state is R under ST column at time = 15:08:14.

Figure 12. Status of the job using condor_q. The job is running.

Figure 13 shows status of the job as completed. We can see state is C under ST column at time = 15:08:20.

Figure 13. Status of the job using condor_q. The job has completed.

In Figure 14, we can see the job gets removed from queue of condor_q. We can see output at time = 15:08:24.

Figure 14. Status of the job using condor_q. The entry of job is removed.

## VII. MAJOR PROBLEMS AND TROUBLE-SHOOTINGS

**Problem 1 and Trouble-shooting**: Not using matching software component depending upon Operating System. Major Grid computing software are targeted for Unix or Linux derivative Operating Systems. Different Linux distributions might use different binary libraries to carry out certain activities. Software related to Grid computing are available in source-code form. If we use source-code and prepare binary files for target machine by compiling them, then there is less probability of having mismatch among various software components. However, if we want to use binary versions, but binary version of the software is not available, then, we try to use software targeted for other matching Linux distributions. Such problem generally occurs between software for Ubuntu and software for Debian.

**Problem 2 and Trouble-shooting**: Trying to run a job without creating proxy credentials. Grid computing uses X.509 certificate based authentication, it does not use authentication through user name and password. When a user wants to use any remote resource for job execution, then the user has to provide its identity to the resource. Therefore, user must create proxy credential to allow successful authentication for the submitted job. Before doing any job submission activity, user can check whether proxy credentials are created or not using grid-proxy-info command.

**Problem 3 and Trouble-shooting**: Certificate error. Certificate error occurs when the clock-time of submitting machine is ahead of the clock-time of target machine. This error reads as "You have sent a certificate with future date/time", as certificate always include validity period, including starting date-time and ending date-time. This problem occurs due to mismatch in clock timings of machines that are present in the Grid computing environment. In our test environment, we alleviate this timing problem by using Network Time Protocol, which does synchronization of clocks with reference server, in our test environment it is grid-b machine.

**Problem 4 and Trouble-shooting**: Unable to understand problem occurring in working of some functionality. We must understand internal working of particular functionality. For example, when we try to test submission and execution of a remote job, nothing happens, i.e., command hangs. As Grid computing environment involves computer network, there might be some delay in getting output of remote job execution. However, if you do not see any output after much longer period than expected, then there is possibility that some internal software component is not working properly. The important thing is we must understand whether problem has occurred due to error in LRM software or error in Grid software.

**Problem 5 and Trouble-shooting**: Unable to understand cause of the problem. Once the location of any problem, either LRM or Grid, is known, the next thing we must understand is what the cause of the error is. If we detect problem due to error in working of LRM software, we must check log files related to LRM and try to understand the problem. Cause of the most of the problems can be known by checking log files and understanding the point of failure.

## VIII. CONCLUSIONS

We related concepts of Cluster computing and Grid computing for the common usage: remote job execution and discussed importance of Grid computing to break barrier of organization boundary, which is existing in Cluster computing. Apart from physical, real Grid computing environment, we also discussed about preparing a virtualized Grid environment using Oracle VirtualBox. To prepare virtual Grid environment, there is no need of many physical machines and networking device. We experimentally tested remote job submission and execution on Condor LRM in it proprietary way, Condor's way. We experimentally tested remote job submission and execution on LRM in Grid computing ways. Accessing resources using Grid computing ways are independent of underlying LRMs.

In future, we would like to explore adding higher-level Grid services and testing of running applications. At present, we have plan to build workflow scheduling and execution environment to allow running of scientific workflows. Moreover, we would also like to test other LRMs such as SGE/OGE and Torque.